\documentclass[prl,aps,twocolumn]{revtex4}
\usepackage{amsmath}
\usepackage{amssymb}
\usepackage {graphicx}
\usepackage{color}
\def\blue{\color{blue}}
\def\red{\color{red}}

\newcommand{\beq}{\begin{equation}}
\newcommand{\eeq}{\end{equation}}
\newcommand{\bea}{\begin{eqnarray}}
\newcommand{\eea}{\end{eqnarray}}
\newcommand{\e}{\varepsilon}
\newcommand{\bk}{{\vec k}}

\newcommand{\bq}{{\vec q}}

\begin{document}

\title{Probing the pairing interaction and multiple Bardasis-Schrieffer modes using Raman spectroscopy}

\author{ S. Maiti $^1$, T.A. Maier$^{2}$, T. B\"{o}hm$^{3,4}$, R. Hackl$^4$ and P. J. Hirschfeld$^1$}
\affiliation{$^1$Department of Physics, University of Florida, Gainesville,
FL 32611}
\affiliation{$^2$Center for Nanophase Materials Sciences and Computer Science and Mathematics Division, Oak Ridge National Laboratory, Oak Ridge, TN 37831-6494, USA}
\affiliation{$^3$Fakult\"at f\"ur Physik E23, Technische Universit\"at M\"unchen, 85748 Garching, Germany}
\affiliation{$^4$Walther Meissner Institute, Bavarian Academy of Sciences and Humanities, 85748 Garching, Germany}

\date{\today}
\begin{abstract}
In  unconventional superconductors, understanding the form of the pairing interaction is the primary goal.  In this regard, Raman spectroscopy is a very useful tool, as it identifies the ground state and also the subleading pairing channels by probing collective modes. Here we propose a general theory for multiband Raman response and identify new features in the spectrum that can provide a robust test for a pairing theory. We identify \emph{multiple} Bardasis-Schrieffer type collective modes and connect the weights of these modes to the sub-leading gap structures within  a microscopic pairing theory. The conclusions are completely general, and we apply our approach to interpret the  B$_{1g}$ Raman scattering in hole-doped BaFe$_2$As$_2$.
\end{abstract}
\maketitle
\emph{Introduction:} Fe-based superconductors (FeSC) appear to display $s$-wave pairing, with an order parameter that may change sign between Fermi surface (FS) pockets\cite{HirschfeldCRAS,ChubukovHirschfeldPhysToday,Qimiao_Review2016,KurokiHosonoReview15}.
{Theoretical calculations based on  spin fluctuations}
have found that the $d$-wave channel can be strongly competitive, and even argued that $d$-wave could become the ground state for sufficiently strong electron\cite{Kuroki08,Graser09} or hole doping\cite{Thomale}. The consequences of a competing pairing channel  were explored by  Bardasis and Schrieffer\cite{BardasisSchrieffer61}, who predicted the existence of a new collective mode corresponding to the phase fluctuations of the subdominant ($d-$wave) order parameter above the ground state  ($s-$wave).
{An analogous simple} calculation was performed by Devereaux and Scalapino\cite{DevereauxScalapino09} for a typical FeSC electronic structure with $s\pm$ symmetry of the ground state with anisotropic gaps. They showed that the mode frequency should depend on $1/u_d-1/\tilde u_s$, where $u_d$ is the $d$-wave coupling constant and $\tilde u_s$ is a renormalized $s$-wave coupling that depends on the angular form of the gap in the condensed state.

Such a mode (called a Bardasis-Schrieffer (BS) mode or particle-particle exciton),   couples to the Raman probe, but was never observed in conventional superconductors.  Recently, however, measurements on Ba$_{1-x}$K$_x$Fe$_2$As$_2$\cite{Hackl12,Hackl14}, NaFe$_{1-x}$Co$_x$As\cite{Blumberg14}, Ba(Fe$_x$Co$_{1-x}$)$_2$As$_2$\cite{CoBa122} found peaks in the B$_{1g}$ polarization spectrum which were consistent with a {collective} mode. Although, {in the latter two cases, these peaks were identified with an excitonic mode originating due to the proximity to the nematic phase\cite{nematicres}, the observation of multiple peaks in the former system for dopings farther from the region dominated by nematic fluctuations is rather puzzling. We thus propose that  the peaks observed in the K-doped system are more likely BS modes.}

In this Letter, we provide a scheme to calculate the Raman-response within a microscopic pairing theory, { using the same microscopic interactions that lead to pairing}, and point out several details of the Raman-spectrum that were either not expected or not explained before. In particular, we show that in crystalline systems multiple BS modes exist and appear with a characteristic weight in the spectrum. We believe that such a theory is necessary to accurately calculate the observed Raman intensity.

In fact, every subleading pairing channel leads to a BS mode ({there may be multiple resonances}). In crystalline systems with unconventional electronic structure, there can be an interplay of several orthogonal form-factors in the subleading channel as well.  This aspect, which can strongly influence the shape of the Raman spectrum, appears to have been neglected until now. Theoretically, the solution to the linearized gap equation for a variety of materials\cite{HirschfeldCRAS} indeed shows the relevance of more than one subleading eigenfunction within the \emph{same} irreducible representation of the normal state symmetry group. In particular, a spin fluctuation pairing calculation for the Ba$_{1-x}$K$_x$Fe$_2$As$_2$ system indicates that the system condenses in a  (A$_{\rm 1g}$) $s_\pm$ state, and that at least two subleading B$_{\rm 1g}$ harmonics have non-negligible eigenvalues. We will show that this situation allows for the existence of well resolved BS peaks in the spectrum whose spectral weight distribution, as obtained from theory, seems consistent with the experiment.

Here we present a general scheme to compute the Raman scattering intensity for a multiband model, including vertex corrections. We then apply the general formulation to two specific cases and illustrate all of the points discussed above. The advantage of this formalism lies in the fact that it is valid for any ground-state with any number of bands and it accounts for all collective modes through the vertex corrections, { which removes the singularity of the Raman response at twice the gap edge.}
\begin{figure}[htp]
\includegraphics[width=1.0\columnwidth]{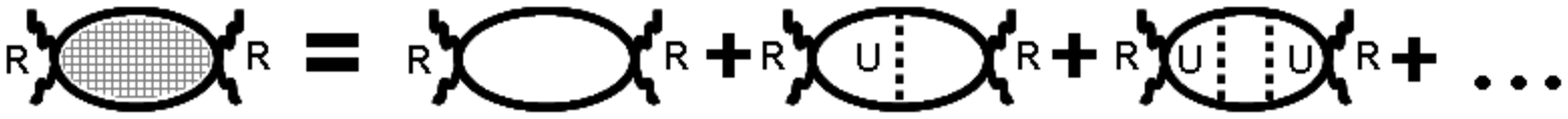}
\caption{
\label{fig:1} The summation scheme for non-resonant Raman-scattering for the B$_{\rm 1g}$ sector. The long-range Coulomb interaction does not affect this sector. Here $U$ is shorthand for a generic residual interaction vertex in the pairing channel.}
\end{figure}
\begin{figure}[htp]
$\begin{array}{c}
\includegraphics[width=\columnwidth]{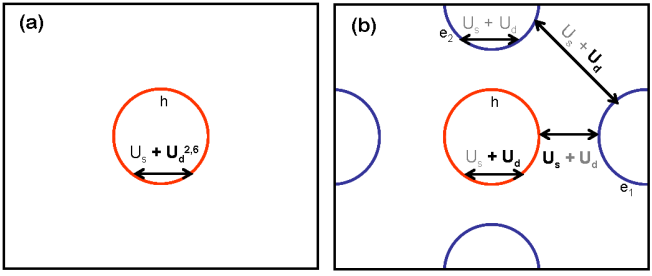}
\end{array}$
\caption{
\label{fig:2} Two toy models considered for illustration of the main results of this work: (a) One pocket around $\Gamma$-point with $s-$ and two $d-$interactions corresponding to $2\theta$ and $6\theta$ harmonics. (b) 3 pockets where only the interactions in the dark font are retained.}
\end{figure}

\emph{Multiband Raman response in the $B_{\rm 1g}$ channel:} The intensity of the Raman response in a multiband system, in the non-resonant response limit, in the B$_{\rm 1g}$ channel can be expressed as\cite{Rev} $\chi_{\rm R}(Q)\equiv\chi_{\rm R}(\Omega,\bq)\propto\sum_{a,b}\int dt e^{-i\Omega t}\langle\rho^{R}_a(t,\bq)\rho^{R}_b(0,\bq) \rangle\equiv \Pi^{RR}(Q)$, where $\rho^R_a$, in the non-resonant limit, is well approximated by the ``Raman density" in the B$_{\rm 1g}$ channel: { $\sum_{\bk}\gamma^a_{\bk}c^{a\,\dag}_{\bk}c^{a}_{\bk},$ where $\gamma^a_{\bk}$ for band $a$ in the B$_{\rm 1g}$ channel is $\partial^2\e^a_{\bk}/\partial k_x^2-\partial^2\e^a_{\bk}/\partial k_y^2$}. Here $\Omega$ and $\bq$ correspond to the shift in frequency and wavevector of the incident light. In metals, $q\ll k_F$, the Fermi wave vector, and will be set to zero in this work. We evaluate the above expression in the SC state using the summation scheme outlined in Fig. \ref{fig:1}. To do so, we work with the Hamiltonian $H=H_0+H_{\rm int}$, where $H_0=\sum_{\bk,\sigma,a}\e^a_{\bk}c^{\dag}_{\bk,\sigma,a} c_{\bk,\sigma,a} - \sum_{\bk}\Delta^*_{\bk,a}c_{\bk,\uparrow,a}c_{-\bk,\downarrow,a} +c.c$ and
\bea
 H_{\rm int}&=&{\small \sum_{a\neq b} U^{(3)}_{ab}(\bq)c^{\dag}_{\bk, \alpha,a}c^{\dag}_{\bk'+\bq,\beta,a}c_{\bk',\beta,b}c_{\bk+\bq,\alpha,b}}\nonumber~~~~~~ \\ &+& \sum_{a} U^{(4)}_{aa}(\bq)c^{\dag}_{\bk,\alpha,a}c^{\dag}_{\bk'+\bq,\beta,a}c_{\bk',\beta,a}c_{\bk+\bq,\alpha,a}.\nonumber
\eea
This is
the momentum dependent form of the interactions as modeled, e.g. in Ref. \cite{ChubukovPhysica,MaitiChubukovRG} (the other interactions neglected here do not affect the main message of this work). We proceed by rotating the basis to the Nambu space with the spinor {$\psi^{\dag}_{\bk}=(c^{\dag}_{\bk,\uparrow,1},c_{-\bk,\downarrow,1}, c^{\dag}_{\bk,\uparrow,2},c_{-\bk,\downarrow,2},...)$},  where $1,2...$ are the various bands. The interaction is recast as {$H_{\rm int}=\sum U^{\alpha\beta\gamma\delta}(\bq)\psi^{\dag}_{\bk,\alpha}
\psi^{\dag}_{\bk'+\bq,\beta}\psi_{\bk',\gamma}\psi_{\bk+\bq,\delta}$.} The explicit form of the interaction vertex is listed in the supplementary material(SM). Then, we need to evaluate
\beq\label{eq:raman2}
\Pi_{RR}(Q)=\sum_{a,b}\int_K \text{Tr}\left[\hat{R}_{a3}\hat{G}_K\hat{\Gamma}^{R_b}_3\hat{G}_{K+Q}\right],
\eeq
\beq\label{eq:raman2s}
\hat{\Gamma}^{R_{a}}_3=\hat{R}_{a3} - \sum_{c,m,n}U^{nm}_{ac}f^n_{\bk}\int_{K'}f^{m*}_{\bk'}
\mathcal{M}_{ac}\cdot\hat{G}_{K'}\hat{\Gamma}^{R_{c}}_3
\hat{G}_{K'+Q}\mathcal{M}^{\dag}_{ac},\nonumber\eeq
where, $\hat{R}_{a3}=\sum_{t}f^t_{\bk}c^t_a[\sigma_3\otimes s_a]$; $\int_K\equiv T\sum_n\int\frac{d^2k}{(2\pi)^2}$; $s_a$ is the band selector of the form diag $(0,...,1,0,...)$ ($1$ at the $a^{\rm th}$ location); $\sigma_{1,2,3}$ are Pauli matrices in  Nambu space; $\hat{G}$ is the bare Greens' function in Nambu space  with elements $\hat{G}_a = [i\omega_n\sigma_0-\epsilon_{\vec{k}}^a\sigma_3-\Delta_{\vec{k},a}\sigma_1]^{-1}$ for band $a$; $\gamma^a_{\bk}$ is expanded as $\sum_t f^t_{\bk} c^t_a$, where $\{f_{\bk}\}$ is a set of orthogonal functions within the B$_{\rm 1g}$ sector; $U^{nm}_{ab}$ is the projection matrix element {of $U^{\alpha\beta\gamma\delta}$} for harmonics $n,m$ for the interaction between bands $a$ and $b$ and $\mathcal{M}$ is a matrix that accounts for transformation to Nambu space (see SM). All matrices are $2N\times2N$, where $N$ is the number of bands and $2$ is the Nambu space dimension. To proceed, it is necessary to introduce the other Nambu components: $\Gamma^{R_b}_i$ with $i=1,2$ in addition to $i=3$, {the solution for which is constructed as $\Gamma^{R_{b}}_{i}=\sum_{t,a}f^t_{\bk}[\sigma_j\otimes s_a]\mathcal{K}^{t,ab}_{ji}$, where the matrix $\mathcal{K}^{t,ab}_{ji}$ is found after substituting for $\Gamma_i^{R_b}$ in Eq. \ref{eq:raman2}. The response $\Pi_{RR}$ is then given by
\bea\label{eq:ra}
\Pi_{RR}&=&\sum_{a,d,t,t'}c^{t,a}\Pi^{tt';a}_{3i}\mathcal{K}^{t',ad}_{i3}, ~\text{where}
\eea
\beq
\mathcal{K}=\left[1+\frac14[U_{pp}]\cdot[\tilde\Pi-\Pi]+\frac14[U_{ph}]\cdot[\tilde\Pi+\Pi]\right]^{-1}
[c],\nonumber\eeq
\beq\label{eq:raman4ab}
\tilde{\Pi}^{m t;b}_{ij}=
\int_{K'}f^{m*}_{\bk'}f^t_{\bk'}\text{Tr}\left[\mathcal{M}_{ab}^{\dag}\cdot[\sigma_i]^a\mathcal{M}_{ab}\hat G_{K'}[\sigma_j]^b \hat G_{K'+Q}\right].
\eeq
Here
$[U_{pp,ph}]$ is the coupling matrix in nambu$\otimes$band$\otimes$harmonic space and $\mathcal{K}$ and $[c]$ are matrices in nambu$\otimes$band space, but a vector in harmonic space  (see SM for examples); The subscript $pp$ and $ph$ for $[U]$ stand for its pairing and density channel projections. Further, $[\sigma_i]^a\equiv \sigma_i\otimes s_a$ and $\Pi$ is the same as $\tilde\Pi$ but without the $\mathcal{M}-\mathcal{M}$ matrices. }

The collective modes are contained in the poles of $\mathcal{K}$.  While in general $\mathcal{K}$ is $4\times4$ in Nambu space, the singlet ground state which preserves time reversal symmetry decouples this into two $2\times2$ blocks: spin-amplitude and phase-density sectors. Since the BS modes are in the phase sector, we only deal with this $2\times2$ subspace in this work. The advantage of this formalism is that the Raman response is computed for a microscopic model where it is dressed by the same interactions that led to pairing. The microscopic problem provides the relevant number of harmonics $H$ that effect the pairing problem and this approach to calculate Raman response then calls for computing the numbers $\Pi^{nm}$ ($n,m\in\{1,...H\}$) and carrying out a matrix inversion.

\emph{Connection to collective modes:} It is well known that the collective modes in the sub-leading channel couple to the Raman response \cite{zawadowskii,DevereauxScalapino09}. However, several questions remain: Are the poles in the  Raman response always the same as the frequencies of collective modes? If there are multiple collective modes, how and with what spectral weight do they couple to the Raman probe?

Our formalism naturally provides answers to such questions. {The poles of Raman response (Eq. \ref{eq:ra}) are contained in  poles of $\mathcal{K}=[1+[U_{pp}][\tilde{\Pi}-\Pi]/4+[U_{ph}][\tilde{\Pi}+\Pi]/4]]^{-1}$. The collective mode in a general multiband superconductor, can be found, e.g. using the formalism in Ref. \cite{maiti15} and they are the poles of $\left[1-[U][\Pi]/2\right]^{-1}$. Thus it is clear that, in general, the poles are \emph{not} the same. However, if the interaction in the density channel is weak ($[U_{ph}]\rightarrow0$), then one can show that in the pairing interaction sector, $\tilde\Pi\rightarrow -\Pi$ and we restore the collective mode result. This was also pointed out in Ref. \cite{Khodas14} using a different scheme.} In most works in the literature, the density channel has been neglected; {for demonstration purposes,} we shall do the same here. This precludes the appearance of a particle-hole exciton\cite{Khodas14} in the Raman intensity analogous to the so-called ``neutron resonance" and does not affect any of our claims for the case of the B$_{1g}$ Raman polarization. The answer regarding the weights of the various BS modes will be apparent in the following examples.

\emph{Simple toy models showing multiple BS modes:} First we consider a $\Gamma$ centered pocket, together  with an $s$-wave BCS like ground state for the system with the order parameter $\Delta$ \cite{FN}. We then choose a competing SC B$_{1g}$ channel with two harmonics via the interaction: $U_{\theta,\theta'}= 2U_{22}\cos2\theta\cos2\theta' + 2U_{66}\cos6\theta\cos6\theta' + 2U_{26}(\cos2\theta\cos6\theta'+\cos6\theta\cos2\theta')$. The gap structure has the form $\Delta_{\theta}=\sqrt{2}\Delta_2\cos2\theta + \sqrt{2}\Delta_6\cos6\theta$. The numbers we need to compute the B$_{1g}$ Raman response are: $\Pi^{22,26,66}_{22,32}$, which in this model are: $\Pi^{22,66}_{22,32}=\Pi^{00}_{22,32}\equiv \Pi_{22,32}$ and $\Pi^{26}_{ij}=0$. We also assume the harmonic decomposition of $\gamma_{\bk}$ in terms of coefficients $c_{2,6}$. The Raman response, as computed in this formalism, is then given by 
\begin{figure*}[htp]
$\begin{array}{ccc}
\includegraphics[width=0.6\columnwidth]{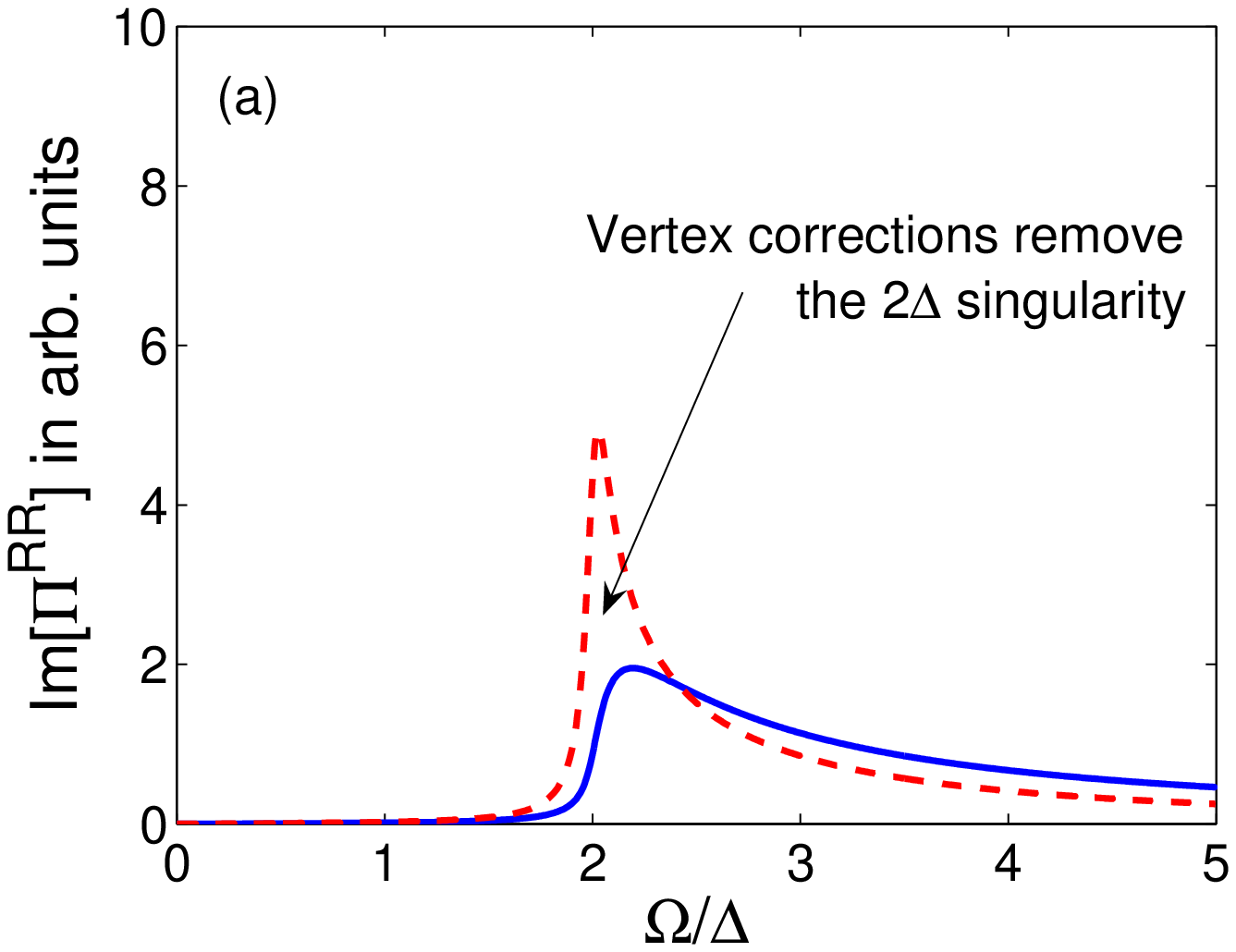}&
\includegraphics[width=0.6\columnwidth]{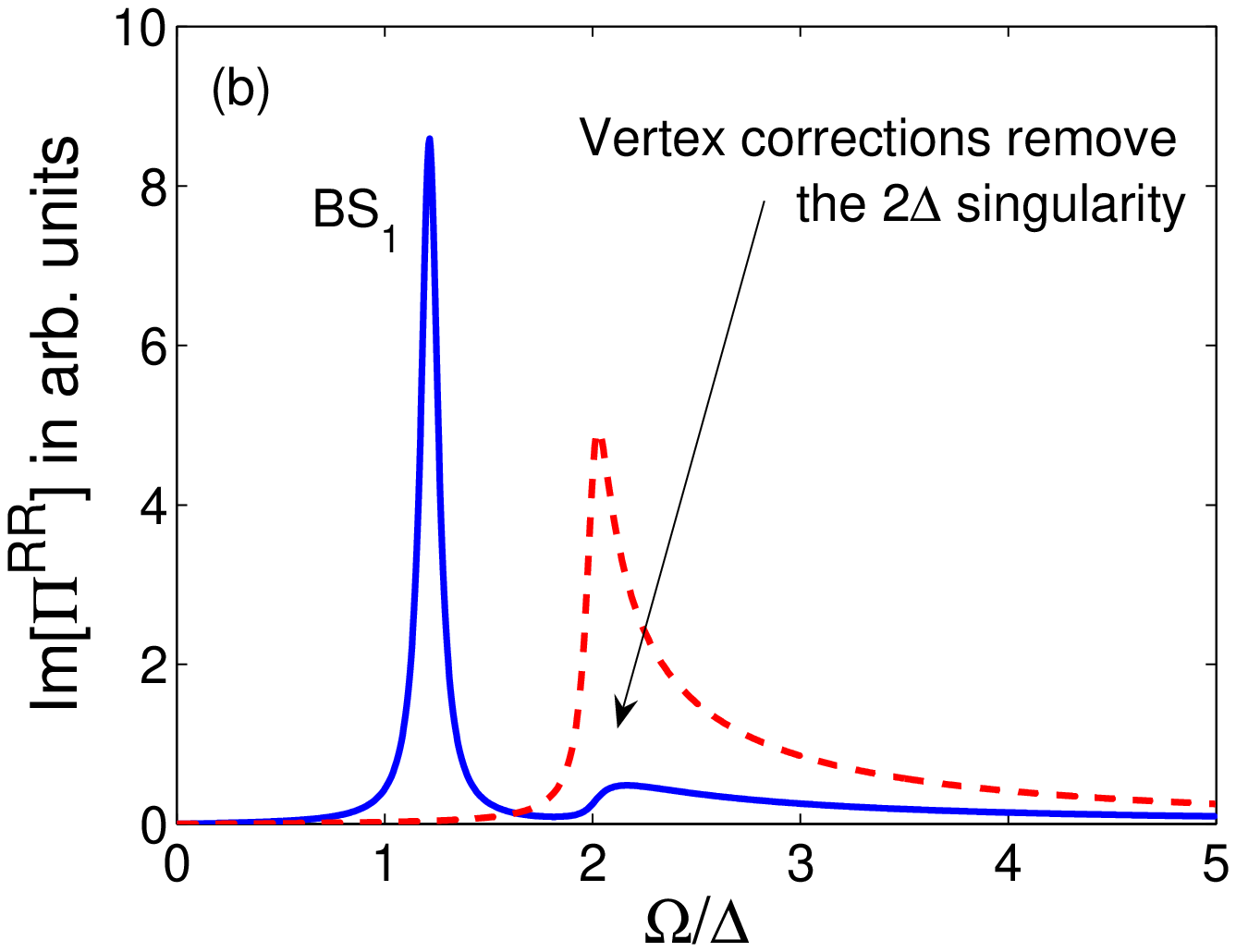}&
\includegraphics[width=0.6\columnwidth]{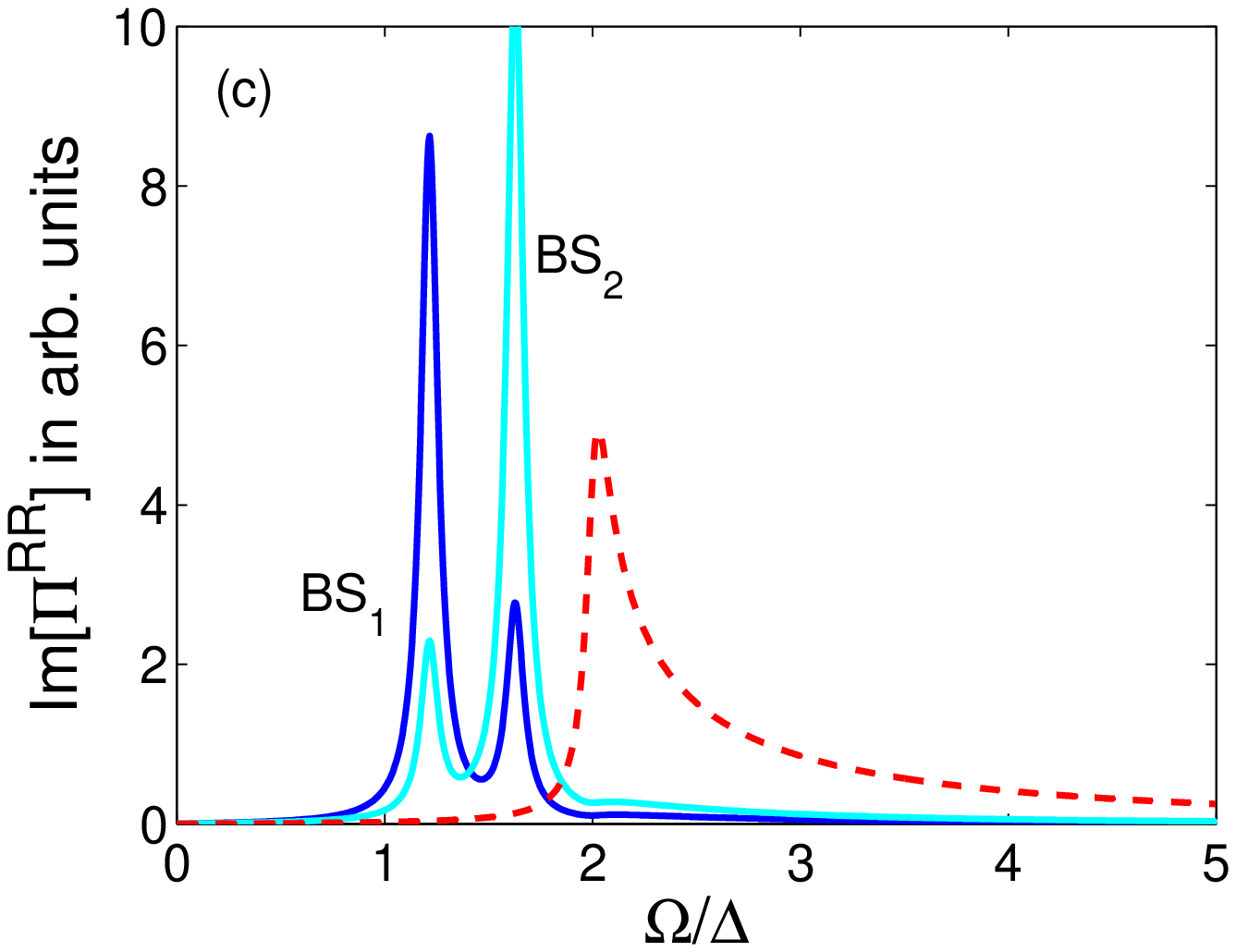}
\end{array}$
\caption{
\label{fig:3} Raman response for toy model with one pocket. The dashed red line in each case is the response in the absence of interactions. (a) Case where $d-$wave is not competitive: $\nu\tilde{U}_1=0.4$ and $\nu\tilde{U}_2=0.3$. (b) Case where only one $d-$wave solution is competitive: $\nu\tilde{U}_{\blue 1}=-0.4$ and $\nu\tilde{U}_2=0.3$. (c) Case where both $d-$wave solutions are competitive: $\nu\tilde{U}_1=-0.4$ and $\nu\tilde{U}_2=-0.3$. The light and dark blue correspond to a band structure such that $\tilde c_{1}$ and $\tilde c_2$ are switched: this shows the connection of spectral weight of a Raman peak with the subleading eigenvectors. Here $\nu U_s=-0.5$, $\tilde c_1=0.6$ and $\tilde c_2=0.3$.  A fermion lifetime of 0.05$\Delta$ was included to get the broadening.}
\end{figure*}
\begin{figure*}[htp]
$\begin{array}{cccc}
\includegraphics[width=0.54\columnwidth]{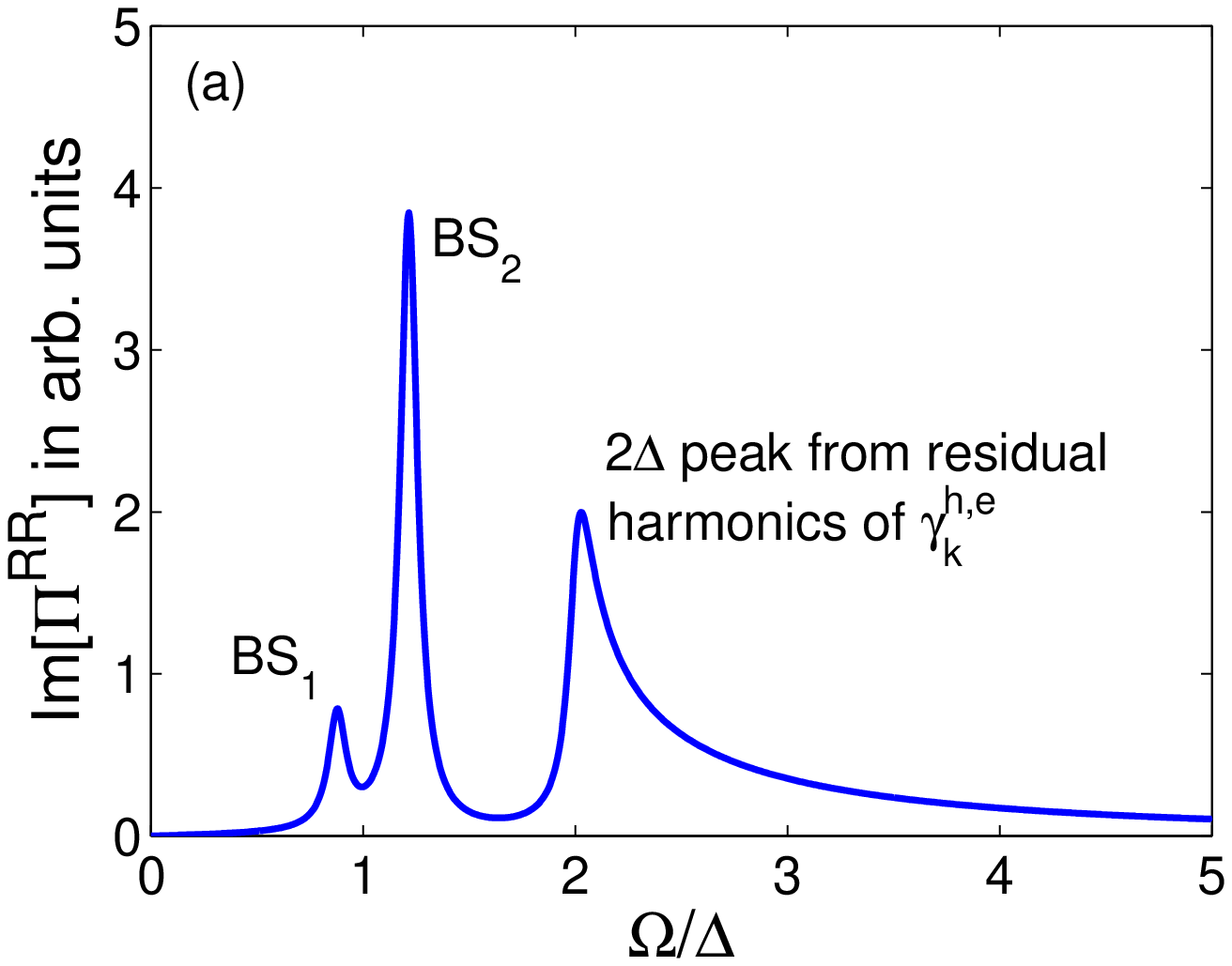}&
\includegraphics[width=0.53\columnwidth]{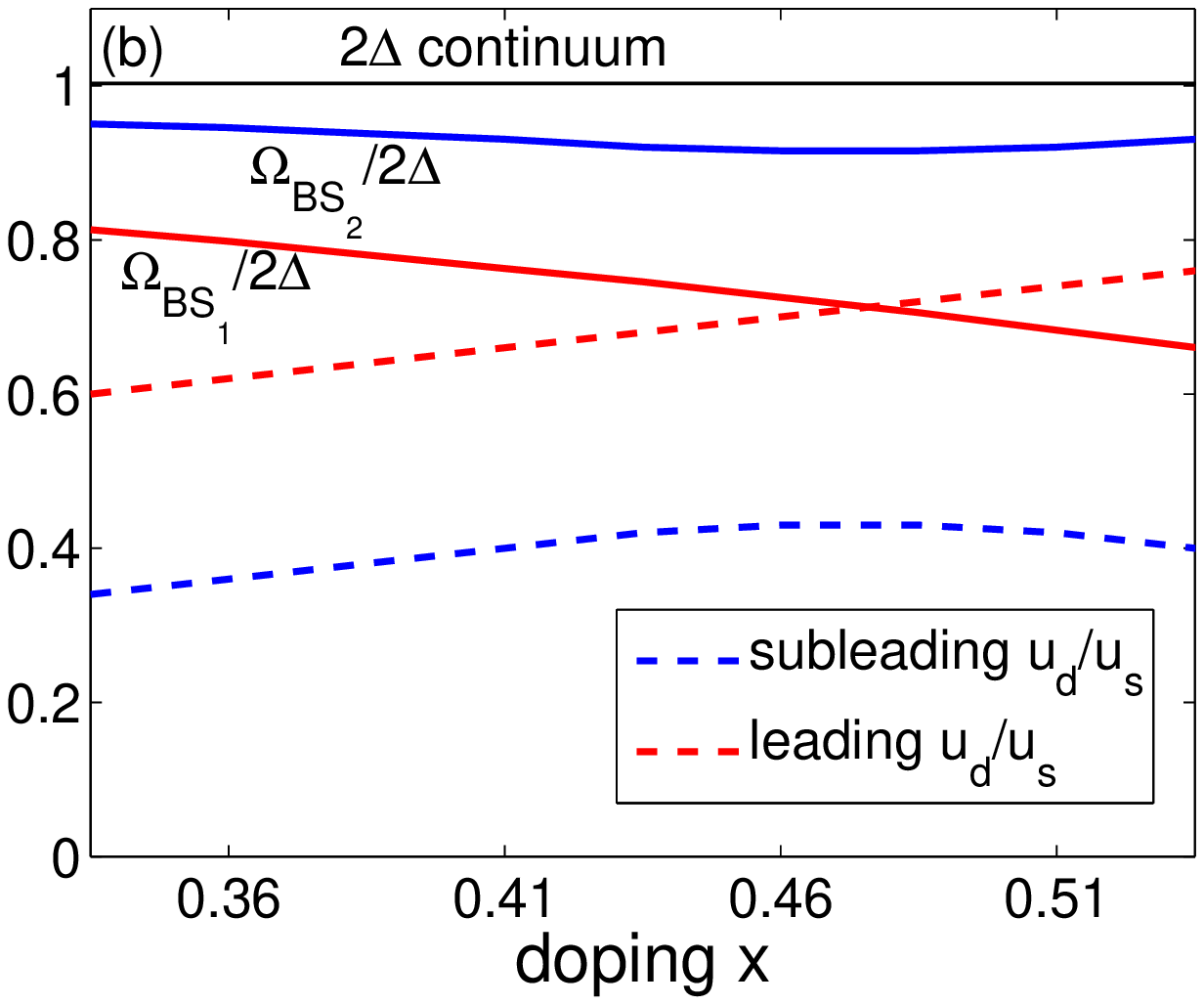}&
\includegraphics[height=1.3in,width=0.43\columnwidth]{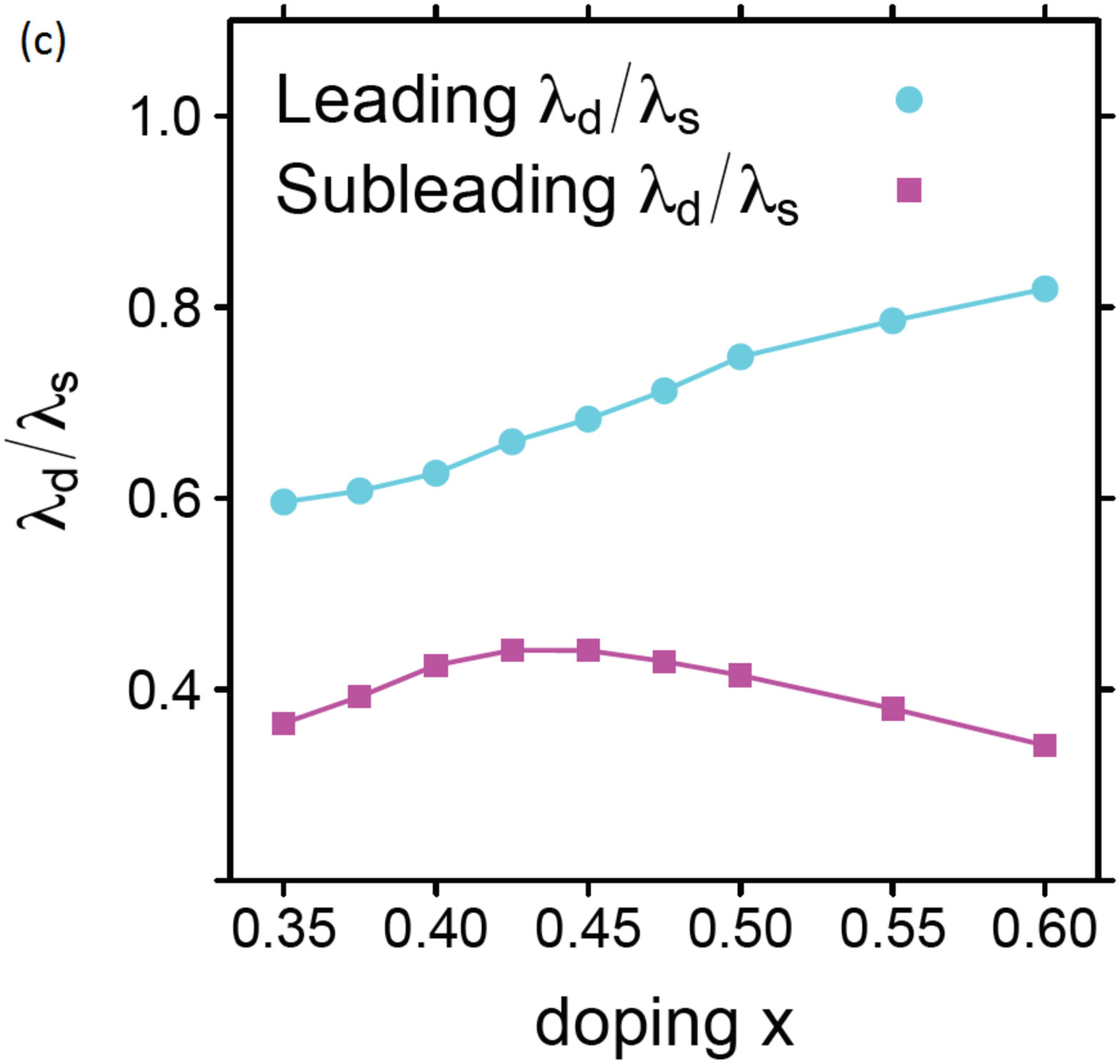}&
\includegraphics[height=1.3in,width=0.45\columnwidth]{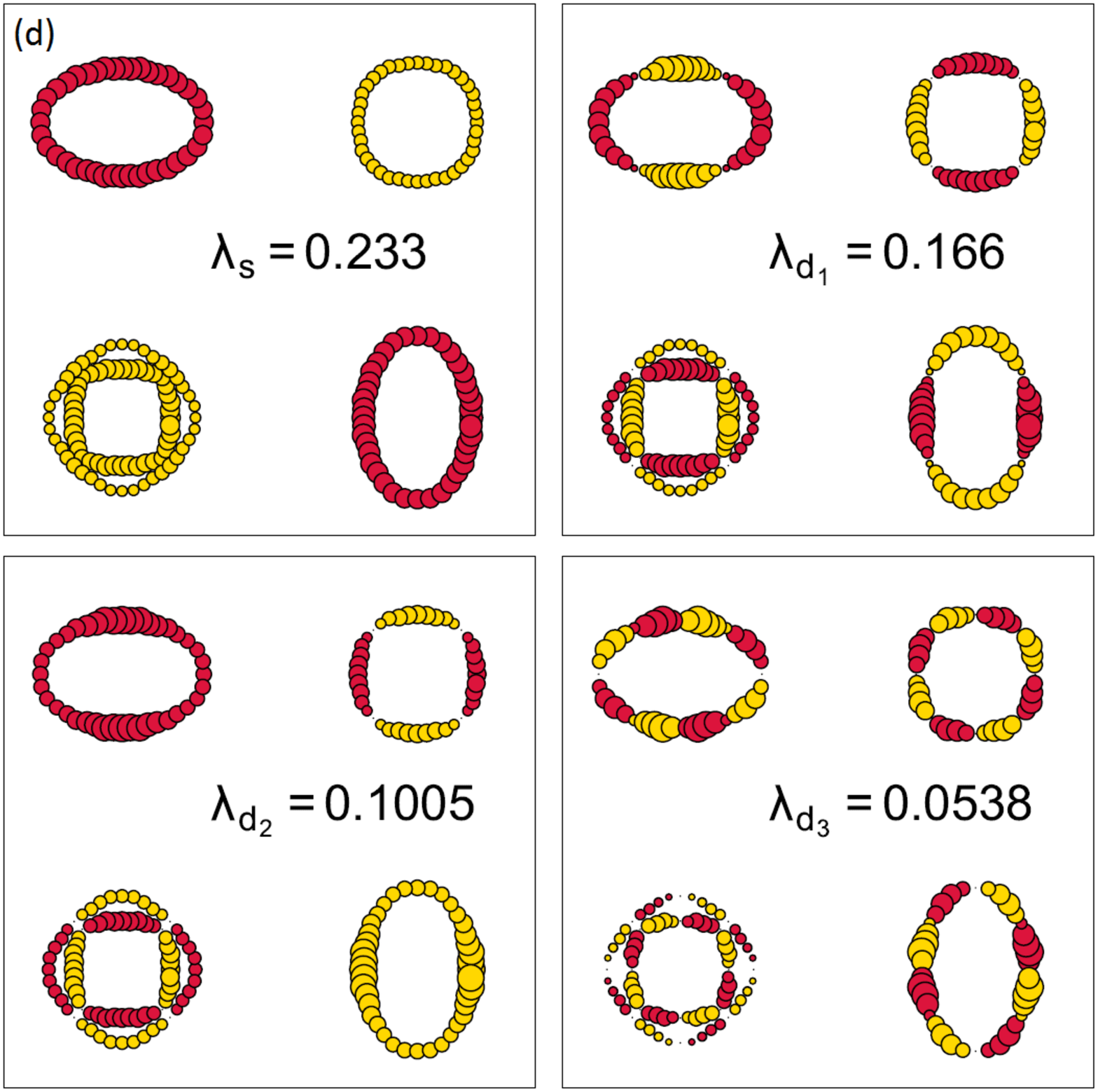}
\end{array}$
\caption{\label{fig:4}
 (a)Raman response in the toy model with 3 pockets,  with parameters $c^h=0.2,~c^e=0.5,~c^o=0.3,~\nu U_{\red s}=0.5$, (broadening of $0.05\Delta$). The $2\Delta$ feature in the spectrum remains because of non-zero $c^o$. (b) Correlating the evolution of the BS peaks by tuning the doping and the leading and subleading $d-$wave eigen-values ($U^{hh,e_1e_2}_d$ are modeled with doping and chosen to mimic panel (c), which is 5-band calculation). (c) Calculated eigenvalues $\lambda_{d_1}$ and $\lambda_{d_2}$ in overdoped Ba$_{1-x}$K$_x$Fe$_2$As$_2$ from RPA (see SM) with doping. (d) Corresponding eigenvectors plotted over the Fermi surface (the $\Gamma-$point has an inner and and outer pocket) for the ground-state $s_\pm$ ($\lambda_{\blue s_1}$), and subsequent $d_{x^2-y^2}$-wave solutions($\lambda_{d_{1,2,3}}$) for $x=0.55$. The symbol size is $\propto$  the gap size with red $=+$ and yellow $=-$.}
\end{figure*}
\begin{widetext}
\beq\label{eq:xxx1}
\Pi^{\rm RR}=(c_2)^2\left\{\Pi_{33}-
\frac{(\Pi_{23})^2\left(U_{22}/2+[U_{26}^2-U_{22}U_{66}]\Pi_{22}/4\right)}
{\mathcal{D}}\right\} +
(c_6)^2\left\{\Pi_{33}-
\frac{(\Pi_{23})^2\left(U_{66}/2+[U_{26}^2-U_{22}U_{66}]\Pi_{22}/4\right)}
{\mathcal{D}}\right\},
\eeq
\end{widetext}
where $\mathcal{D}\equiv 1-(U_{22}+U_{66})\Pi_{22}/2-(\Pi_{22})^2(U^2_{26}-U_{22}U_{66})/4$, { zeroes of which correspond to peaks in the Raman spectrum.} A simple exercise\cite{SM1} shows us that this determinant is exactly the equation the determines the frequency of the BS modes. Important information about the pairing interaction can be more readily extracted if we rotate the interaction in the orthogonal basis functions provided by the eigen vectors of the pairing problem. When this rotation is done,{ $\{U_{22},U_{26},U_{66}\}\rightarrow\{\tilde{U}_1,0,\tilde{U}_2\}$, $\{c_{2},c_{6}\}\rightarrow\{\tilde{c}_1,\tilde{c}_2\}$; where $\tilde{c}_{1,2}=\int_{\theta}\left(\frac{\partial^2\e}{\partial k_x^2}-\frac{\partial^2\e}{\partial k_y^2}\right)\left.\right|_{FS}\Delta^{(1,2)}_{\theta}$ are the overlap of the Raman vertex $\gamma_{\bk}$ with the eigenvectors $\Delta^{(1,2)}_{\theta}$ (FS stands for projection on the Fermi surface). The response then takes the form $\Pi^{RR}=$
\beq\label{eq:x}
(\tilde{c}_1)^2\left\{\Pi_{33}-
\frac{(\Pi_{23})^2}
{\frac{2}{\tilde{U}_1}-\Pi_{22}}\right\}  +(\tilde{c}_2)^2\left\{\Pi_{33}-\frac{(\Pi_{23})^2}
{\frac{2}{\tilde{U}_2}-\Pi_{22}}\right\}.
\eeq
Here $\Pi_{22}=\frac{2}{U_s}-2\nu F(\Omega)$,
where $\nu$ is the density of states at the FS and  { $F(\Omega)= \frac{\left(\Omega/2\Delta\right)\sin^{-1}\left(\Omega/2\Delta\right)}{ \sqrt{1-\left(\Omega/2\Delta\right)^2}}$}. The BS modes are solutions to $\nu F(\Omega)=-\frac{1}{\tilde{U}_{1,2}}+\frac{1}{U_s}$. These BS modes are weighted by $\tilde{c}_{1,2}$. {Note that the weight of the BS mode goes to zero as it softens\cite{SM2}.} Fig. \ref{fig:3} displays the Raman response for various cases: when $d-$wave solution is not competing ($\tilde{U}_{1,2}>0$, i.e. repulsive channel), there are no collective modes. As more channels become competitive, collective modes begin to show up (Fig \ref{fig:3}b-c). The weight of the collective modes are controlled by the electronic structure via the $\tilde{c}_1$ and $\tilde{c}_2$ coefficients. Since a microscopic theory for pairing is capable of providing the numbers $\tilde{U}_{1,2},~\tilde{c}_{1,2}$, augmenting such a theory with a calculation of Raman spectrum provides a much stronger testing ground for its validity. It is also clear from Fig. \ref{fig:3} that the harmonics of the interaction that `host' the BS modes contribute very little to the $2\Delta$ peak, an effect due to vertex correction and is analogous to what happens in A$_{\rm 1g}$ sector \cite{Lara}. There are contributions, however, to the $2\Delta$ peak from other solutions that have an eigenvalue close to zero and the effect of vertex corrections is weak. This is shown explicitly in the next example.

 We now consider an example where the electronic structure is more specific to FeSC. Such a model would consist of 1 hole and 2 electron pockets as shown in Fig. \ref{fig:2}. To minimize the parameters and keep the calculations analytically tractable, we choose  minimal interactions necessary to satisfy the symmetry requirement for the ground state  of FeSC:  we first restrict  the  interband interaction to be $U_s$ in the $s$-channel, and further assume the following relations for the density of states at the Fermi level: $\nu_h=2\nu_e\equiv\nu$. This results in an $s\pm$ state with $\Delta_h=-\Delta_e=\Delta$.  We now choose the interaction in the B$_{\rm 1g}$ channel with only $U^{hh}_d$ and $U^{e_1e_2}_d$ components retained.  This guarantees two competing subleading solutions: $[\Delta_h=\cos2\theta;~\Delta_{e_1,e_2}=0]$ and $[\Delta_h=0;~\Delta_{e_1,e_2}=(1,-1)]$.  We now define the overlaps of $\gamma^{h,e}(\bk)$ with harmonics in the interaction above to be $c^h$ and $c^e$. The  overlaps with remaining harmonics not dominant in the interaction are lumped under $c^o$. Following the same procedure as above and rotating the B$_{1g}$ interaction in the pairing eigenvector basis, we find the Raman response to be analogous to Eq. \ref{eq:x} with $\tilde c_{1,2}\rightarrow c^{h,e}$ and an additional term, $(c^o)^2\{\Pi^{h}_{33}+\Pi^{e}_{33}\}$, from the residual harmonic content of $\gamma^{h,e}_{\bk}$. This latter term is responsible for the  $2\Delta$ peak in the presence of BS modes, and represents the combined weight of all the Bardasis-Schrieffer modes of negligible strength, piled up around the two-particle continuum edge.  In previous calculations,   these contributions were neglected, so, if the  collective modes were properly accounted for, the   $2\Delta$ peak was absent, in contrast to  experiments\cite{Hackl12,Hackl14,Blumberg14}. This is the first explanation, to our knowledge, of  this essential experimental feature.  While the straightforward algebra is shown in SM, the Raman response is plotted and explained in Fig.\ref{fig:4}.

\emph{Relevance to (Ba,K)Fe$_2$As$_2$:} We now wish to apply this new {understanding of the Raman spectrum} to the overdoped region of BaKFe$_2$As$_2$. It is well known \cite{Maiti11a,Maiti11b,Thomale} in this system that higher hole doping makes the $d$-wave state competitive with the $s\pm$ ground state.  We have carried out RPA calculations for the 5-orbital model for BaFe$_2$As$_2$ introduced in Graser et al.\cite{Graser10} using the usual spin- and charge-exchange interaction\cite{bickers89,takimoto04}. As shown in Fig. 4c and d, these calculations find a leading $s\pm$ state and at least two subleading and competing B$_{\rm 1g}$ states which are well resolved in energy\cite{FN2}. Consequently, the insight from this work suggests appearance of two BS modes as a function of doping. Such a feature is reportedly seen in experiments\cite{PC} }where the trend in the evolution of the peak positions { correlates with} the trend in eigenvalues just as shown for the toy model {in Fig.\ref{fig:4}b}. 

\emph{Conclusions:}
In summary, we have provided a proof of principle method for using the details of the Raman spectrum, together with theory, to {learn} the details of the pairing interaction in an unconventional superconductor. The calculation of the response with  the full momentum structure of the interaction {is outside the scope of this letter} and will be considered in a more detailed future study. This formalism  is readily generalizable to any ground-state symmetry and any number of bands.  We have identified several features that help to better understand the Raman spectrum: a)  there can be multiple Bardasis-Schrieffer modes in an $s$-wave superconductor; b) {the overlap of the gap structure with the bare Raman vertex $\gamma_{\bk}$ determines the weights of the modes} c)  incorporating the vertex corrections, we find that the $2\Delta$ feature {is suppressed and {exists  only due to} the residual harmonics of the Raman vertex $\gamma_{\bk}$ that are not involved in pairing.}

\emph{Acknowledgments:} The authors are grateful for useful discussions with L. Benfatto, A. Chubukov, D. Einzel, and D. Scalapino.  PJH was supported by US Department of Energy grant DE-FG02-05ER46236. The RPA calculations were conducted at the Center for Nanophase Materials Sciences, which is a DOE Office of Science User Facility.  PJH's work was performed in part at the Aspen Center for Physics, which is supported by National Science Foundation grant PHY-1066293.
\newpage
\begin{widetext}
\section*{Supplementary Material}
\subsection{Interactions in Nambu Space and $\mathcal{M}$ matrices}
Here we present the full structure of the interaction terms $U^{\alpha\beta\gamma\delta}$ used in this work. As far as the multiband pairing problem is concerned, $H_{\text{int}}$ has the following terms corresponding to intraband interactions and interband interactions. Momentum conservation (unless there are identically overlapping Fermi-surfaces) leads to only intraband and pairwise interband interactions. For every band $a$, we have interaction terms corresponding to interacting with itself and another band $b$. The complete breakdown is given as:
\bea\label{eq:4}
H_{\text{int}}&=&\sum_{a,b}H^{ab}_{\text{int}}\nonumber\\
H^{ab}_{\text{int}}&=&H^{ab}_3 + H^{aa}_4\nonumber\\
H^{ab}_3&=&\sum_{\bk\bk'\bq}U^{(3)}_{ab}(\bq)
\psi^{\dag}_{\alpha}(\bk)\psi^{\dag}_{\beta}(\bk'+\bq)\psi_{\gamma}(\bk')\psi_{\delta}(\bk+\bq)
\left([P^+_+]_{\alpha\delta}[P^+_{-}]^{\dag}_{\beta\gamma}+[P^-_{-}]_{\alpha\delta}[P^-_+]^{\dag}_{\beta\gamma}\right)\\
H^{aa}_4&=&\sum_{\bk\bk'\bq j}U^{(4)}_{aa}(\bq)
\psi^{\dag}_{\alpha}(\bk)\psi^{\dag}_{\beta}(\bk'+\bq)\psi_{\gamma}(\bk')\psi_{\delta}(\bk+\bq)
[N^a_+]_{\alpha\delta}\cdot[N^a_j]_{\beta\gamma}\\
\text{where}
&&[P^{\pm}_+]_{\alpha\beta}=
\frac14[(\sigma_3+\sigma_0)\otimes(\sigma_1\pm i\sigma_2)]_{\alpha\beta},\nonumber\\
&&[P^{\pm}_{-}]_{\alpha\beta}=
\frac14[(\sigma_3-\sigma_0)\otimes(\sigma_1\pm i\sigma_2)]_{\alpha\beta}.\nonumber\\
&&[N^{a}_{\pm}]_{\alpha\beta}=
\frac12[(\sigma_3\pm\sigma_0)\otimes s_a]_{\alpha\beta},\nonumber
\eea
$j\in(+,-)$, $\sigma_1\pm i\sigma_2$ is understood to act on the $a-b$ subspace. In the above matrices $P,~N$, the inner-most dimension corresponds to the nambu space which is $2\times2$ in our problem due to spin-rotational invariance. The band space is $N\times N$ and thus the interaction vertex is a juxtaposition of two $2N\times2N$ matrices as shown below in an example for the 2 band case. In this example, $U_{3}$ is the interband interaction and $U_{4,5}$ are the intra-band interactions. They take the form:
\beq\label{eq:int_223}
U^{(3)}[P][P]\rightarrow U_3
\left(\begin{array}{cccc}
0&0&1&0\\
0&0&0&0\\
0&0&0&0\\
0&0&0&0\\
\end{array}\right)\left(\begin{array}{cccc}
0&0&0&0\\
0&0&0&0\\
0&0&0&0\\
0&-1&0&0\\
\end{array}\right)+
U_3
\left(\begin{array}{cccc}
0&0&0&0\\
0&0&0&0\\
0&0&0&0\\
0&-1&0&0\\
\end{array}\right)\left(\begin{array}{cccc}
0&0&1&0\\
0&0&0&0\\
0&0&0&0\\
0&0&0&0\\
\end{array}\right);
\eeq
\beq\label{eq:int_223a}
U^{(4)}_{hh}[N][N]\rightarrow U_4
\left(\begin{array}{cccc}
1&0&0&0\\
0&0&0&0\\
0&0&0&0\\
0&0&0&0\\
\end{array}\right)\left(\begin{array}{cccc}
1&0&0&0\\
0&0&0&0\\
0&0&0&0\\
0&0&0&0\\
\end{array}\right)
+U_4
\left(\begin{array}{cccc}
0&0&0&0\\
0&-1&0&0\\
0&0&0&0\\
0&0&0&0\\
\end{array}\right)\left(\begin{array}{cccc}
0&0&0&0\\
0&-1&0&0\\
0&0&0&0\\
0&0&0&0\\
\end{array}\right);
\eeq
\beq\label{eq:int_223b}
U^{(4)}_{ee}[N][N]\rightarrow U_5
\left(\begin{array}{cccc}
0&0&0&0\\
0&0&0&0\\
0&0&1&0\\
0&0&0&0\\
\end{array}\right)\left(\begin{array}{cccc}
0&0&0&0\\
0&0&0&0\\
0&0&1&0\\
0&0&0&0\\
\end{array}\right)
+U_5
\left(\begin{array}{cccc}
0&0&0&0\\
0&0&0&0\\
0&0&0&0\\
0&0&0&-1\\
\end{array}\right)\left(\begin{array}{cccc}
0&0&0&0\\
0&0&0&0\\
0&0&0&0\\
0&0&0&-1\\
\end{array}\right);
\eeq

The notation $U_{ab}\mathcal{M}_{ab}\cdot\mathcal{M}_{ab}$ is short for terms precisely of the type shown in (\ref{eq:int_223}), (\ref{eq:int_223a}), and (\ref{eq:int_223b}). $\mathcal{M}_{ab}$ takes the form of $[P]$ or $[N]$ depending on $U_{aa}$ or $U_{ab}$ ($b\neq a$) and the dot product is short for the implied sum over all the terms as in (\ref{eq:int_223}), (\ref{eq:int_223a}), and (\ref{eq:int_223b}). The generalization to multiple bands is now obvious.

\subsection{The Raman kernel $\mathcal{K}$}
The self consistent equation for $\mathcal{K}$ that we get after plugging in the form of $\Gamma_i^{R_b}$ in to Eq. (1) of the main text is:
\bea\label{eq:raman4a}
\mathcal{K}^{t,gd}_{pi}&=&c^{t}_g\delta_{pi}\delta_{gd}-\sum_{a,b,c,t',m}\frac{U^{tm}_{ab}}{2}
\int_{K'}f^{m*}_{\bk'}f^{t'}_{\bk'}\text{Tr}
\left[\mathcal{M}_{ab}^{\dag}\cdot[\sigma_p\otimes s_g]\mathcal{M}_{ab}\hat G_{K'}[\sigma_j\otimes s_c] \hat{G}_{K'+Q}\right]\mathcal{K}^{t',cd}_{ji}.
\eea
It is useful to note that $\mathcal{M}_{ab}^{\dag} [\sigma_j\otimes s_a]\mathcal{M}_{ab}=[(\sigma_3\pm\sigma_0)\sigma_j(\sigma_3\pm\sigma_0)\otimes s_b]$ and zero otherwise. Thus $a\rightarrow g$ in Eq. \ref{eq:raman4a}. Also due to diagonal-in-band structure of $G$, $c\rightarrow b$. This leads to the matrix form: $\mathcal{K}=\left[1+ \frac{[U]}{2}\cdot[\tilde\Pi] \right]^{-1}$. The meaning of $[U]\cdot[\tilde\Pi]$ is explained below:
\bea\label{eq:godknows}
[U]\cdot[\tilde\Pi]&=&\frac12[U^{3;x}][\tilde\Pi] +\frac12[U^{3;y}][\Pi]+[U^{4}][\tilde\Pi] + [U^{5}][\tilde\Pi].
\eea
where $[U]'s$ are $4\times 4$ matrices ($2N=4$ for this case) given by (each block is a $2\times2$ matrix),
\beq\label{eq:godknows2}
[U^{3;x}]=\left(\begin{array}{cc}
0&U^3\\
U^3&0
\end{array}\right);~~
[U^{3;y}]=\left(\begin{array}{cc}
0&-U^3\\
-U^3&0
\end{array}\right);~~
[U^{4}]=\left(\begin{array}{cc}
U^4&0\\
0&0
\end{array}\right);~~
[U^{5}]=\left(\begin{array}{cc}
0&0\\
0&U^5
\end{array}\right).
\eeq

In Eq. \ref{eq:godknows}, the $U^{4,5}$ terms can be written as $\frac12[U^{4,5}][\tilde\Pi-\Pi] + \frac12[U^{4,5}][\tilde\Pi+\Pi]$. The first term is what usually occurs in the pairing problem. The second terms is the density channel contribution. This is decomposition of an interaction into $U_{pp}$ and $U_{ph}$ components. In the usual `pairing-interaction' approximation, one drops this contribution. Then we finally get
\beq\label{eq:godknows4}
[U]\cdot[\tilde\Pi]=\frac12[U_{pp}][\tilde\Pi-\Pi],
\eeq
where [U] know takes the form of the pairing kernel:
\beq\label{eq:godknows4}
[U_{pp}]=\left(\begin{array}{cc}
U^4&U^3\\
U^3&U^5
\end{array}\right)
\eeq
This is the form presented in the examples in the main text. All the elements are $\bk,\bk'$ dependent. They are then decomposed into harmonics.

\subsection{Multiple Bardasis-Schreiffer modes in a 1-band model}
This is a digression from the main theme of the Letter (which is to compute the Raman response), but we wish to show that multiple BS modes is not a feature of Raman but intrinsic to the superconductor. Thus, in this section, we find the collectives independent of the selectivity effects of the Raman vertex. Since the Bardasis-Schreiffer modes do not couple to charge fluctuations directly, we might as well find the collective modes in the $d-$sector for a neutral superconductor. This can immediately found using the result from Ref. \cite{maiti15}. The mode equation that gives all the collective modes is:
\bea\label{eq:1}
\left\{\Pi^{L_a,L'_b}_{i,j}-2[V^{L_a,L'_b}]^{-1}\delta_{ij}\right\}\delta F^{L'_b}_j &=&0.
\eea
[Repeated indices are summed over]. $\{i,j\}\in{1,2}$ are the Nambu-space indices corresponding to the amplitude (1) and phase(2) sectors. $L,~L'$ are the irreducible representations and their subscripts $a,b$ run over the different orthogonal harmonics supported in that representation. In what follows, we will limit ourselves to the $B_{\text{1g}}$ sector. Because there is no time reversal symmetry breaking, $\Pi_{12}=0$. Thus we will only look at $\Pi_{22}^{nn'}$ form (we stick to $L=B_{\rm 1g}$ an $n$ refers to the harmonics in B$_{1g}$). To start, our $s-$wave state is the one with the 1-band isotropic gap structure with $\Delta_s=\Delta$. For brevity, we shall now use
\beq\label{eq:1.5}
\Pi_{22}^{nn'}(\Omega)\equiv P^{nn'}(\Omega)=\delta_{nn'}\nu_{2D}\left[-4L -\left(\frac{\Omega}{2\Delta}\right)^2I_0(\Omega)\right],~~~L\equiv\frac12\ln\frac{2\Lambda}{\Delta}.
\eeq

The other ingredients for BS modes are $T=T_c$ solutions for the $d-$wave sector and $T=0$ solutions for the $s-$wave sector. For the $d-$sector, the usual procedure results in the following gap equation:
\bea\label{eq:2}
&&\Delta_{\theta}=-\int_{\theta'}U_{\theta,\theta'}\Delta_{\theta'}l,\\
&\Rightarrow&1=-U\int_{\theta'}\left(f^d_{\theta'}\right)^2l,
\eea
where it is assumed that $U_{\theta,\theta'}=U f^d_{\theta}f^d_{\theta'}$, $\Delta_{\theta}=\Delta_d f^d_{\theta}$, and $f^d_{\theta}=c_n\cos n\theta$, $n\in\{2,6,10, ...\}$ (with the normalization $\sum_n c_n^2 = 1$). This clearly gives one $d-$wave solution. This then leads to one solution of the mode equation and thus one BS mode. But now, lets pick the following form of the interaction
\bea\label{eq:3}
U_{\theta,\theta'}&=& U_d\cos2\theta\cos2\theta' + r U_d\cos6\theta\cos6\theta' + V(\cos2\theta\cos6\theta'+\cos6\theta\cos2\theta'),\nonumber\\
\Delta_{\theta}&=&\Delta_2\cos2\theta + \Delta_6\cos6\theta.
\eea
We then get the following matrix equation for $T_c$ ($l\sim\ln\Lambda/T^d_c$: Here $T_c^d$ refers to the $T_c$ that would have resulted in the absence of the $s-$wave channel)
\bea\label{eq:4}
&&\text{det}\left(
\begin{array}{cc}
1+u_d l&vl\\
vl&1+r u_dl
\end{array}\right)=0\nonumber\\
&\Rightarrow&\text{det}[l+[V]^{-1}]
=0\nonumber\\
\text{where,~~~} &&[V]=\left(
\begin{array}{cc}
u_d &v\\
v&r u_d
\end{array}\right)
\eea
Here the smaller case symbols are dimensionless couplings obtained after multiplying the upper case symbols by the DOS $m/2\pi$. There are two possible values of $T_c$ given by ($u_d<0$ for d-channel to be attractive)
\bea\label{eq:5}
l_{\pm}=\frac{-(1+r)u_d\pm\sqrt{(1-r)^2u^2_d+4v^2}}{2\left(ru_d^2-v^2\right)}.
\eea
For the $s-$sector at $T=0$, we get
\beq\label{eq:6}
\frac{1}{u_s}=-2L.
\eeq
The $L$ is the same as in Eq. \ref{eq:1.5}. The mode equation \ref{eq:1}, now reads
\beq\label{eq:7}
\text{det}\left([P]-2[V]^{-1}\right)=0,~~\text{where}~~[P]=\text{diag}(P^{22},P^{66}).
\eeq
In fact, $P^{22}=P^{66}\equiv P$. All other elements of $[P]$ are zero. Note that this equation is the same as the eigen value equation in Eq. \ref{eq:4}. Thus the BS collective modes are given by $P=-2 l_{\pm}$. These roots are exactly the same as the solutions to $\mathcal{D}=0$ in the main text.

\subsection{Evaluation of Raman response for the 3-pocket model}
Here we provide the details of the straightforward algebra to calculate the Raman response for the 3 pocket model described in the main text. At $T=0$ we have the following for the ground state of the system (same notation and parameters in the main text):
\bea\label{eq:lk}
\Delta_h&=&-2U_s\nu_e\Delta_eL_e,~~L_e=\ln\frac{2\Lambda}{|\Delta_e|},\nonumber\\
\Delta_e&=&-U_s\nu_h\Delta_hL_h,~~L_h=\ln\frac{2\Lambda}{|\Delta_h|},
\eea
where $\Lambda$ is a pairing cut-off energy. For the choice of density of states made in the main text, we get the ground state to be $\Delta_h=-\Delta_e=\Delta>0$ and $L_{e,h}=\ln\frac{2\Lambda}{\Delta}$. To get to the B$_{\rm 1g}$ sector, it is necessary to define the harmonics $f^n_{\bk}$. These belong to the following set: $\{\pm1, \cos2\theta, \cos6\theta,... \}$. It is understood that `$\pm1$' is applicable only to off-$\Gamma$ point pockets (electron pockets in this case). In this sense, the $d-$wave gap structure, in the harmonic expansion, takes the form
\bea\label{eq:ggg}
\Delta_h(\theta)&=&0*1 + \Delta^{(2)}_h \cos2\theta+...\nonumber\\
\Delta_{e_1}(\theta)&=&\Delta^{(1)}_e*1 + \Delta^{(2)}_{e} \cos2\theta+...\nonumber\\
\Delta_{e_2}(\theta)&=&-\Delta^{(1)}_{e}*1 + \Delta^{(2)}_{e} \cos2\theta+...\nonumber
\eea
Of course, for the choice of interactions in the main text, $\Delta^{(2)}_{e_1,e_2}=0$. Switching to the notation in the main text $\Delta^{(2)}_h\rightarrow \Delta_h$ and $\Delta^{(1)}_e\rightarrow \Delta_e$, the eigen value set up for the $d-$wave component is
\bea\label{eq:lk1}
\Delta_h&=&-U^{(hh)}_d\nu_h\Delta^{(d)}_h l,~~l=\ln\frac{\tilde\Lambda}{T_c},\nonumber\\
\Delta_{e_1}&=&-U^{(e_1e_2)}_d\nu_e \Delta_{e_2}l, \text{and}\nonumber\\
\Delta_{e_2}&=&-U^{(e_1e_2)}_d\nu_e \Delta_{e_1}l,
\eea
where we understand that symmetry requirements require $\Delta_{e_1}=-\Delta_{e_2}=\Delta_e$. There are two solutions to the above problem: $\Delta_{h}\neq 0, \Delta_{e}=0$ with the $T_c$ given by $l=1/u^{(hh)}_d$ and $\Delta_{h}= 0, \Delta_{e}\neq0$ with the $T_c$ given by $l=2/u^{(e_1e_2)}_d$. We thus expect two BS modes and we will now see how this is shown using the formula in the main text. We need the poles of the $\mathcal{K}$ matrix which involves $[U]$ and $[\tilde\Pi-\Pi]$. We will have only two harmonics in the interaction, but assume that the $\gamma_{\bk}$ matrix requires 3 harmonics (whatever the third harmonic may be). Thus matrix $[U]$ in this problem is:
\beq\label{eq:matr}
[U]=\left(\begin{array}{cccccc}
0&0&0&0&0&0\\
0&-U^{(e_1e_2)}_{2\times2}&0&0&0&0\\
0&0&U^{(hh)}_{2\times2}&0&0&0\\
0&0&0&0&0&0\\
0&0&0&0&0&0\\
0&0&0&0&0&0
\end{array}
\right),
\eeq
where $A_{2\times2}\equiv A 1_{2\times2}$. As a reminder, the space is nambu$\otimes$band$\otimes$harmonic. A note on how to get $[U]$ from pairing $U_{\bk,\bk'}$: We ignore the Nambu space for now ($[U]\propto 1_{2\times2}$ in that space). The other entries  $U_{ab}^{nm}$ (with dimension $N\times n$) correspond to $\int_{\bk}\int_{\bk'}f^n_{\bk'}U^{ab}_{\bk',\bk}f^m_{\bk}$. If $a$ and $b$ are the same bands, then we sum over all pairs of pockets to get $U^{aa}$. For example, the $d-$wave pairing matrix for the electron pockets is
\beq
U_{pp}=\left(\begin{array}{cc}
u&v\\
v&u
\end{array}
\right).
\eeq
The relevant eigenvector is $e^T=(1,-1)/\sqrt{2}$. The $[U]$ matrix is just $1\times1$ because the FS harmonic this problem is using is $\{1\}$ form the set $\{1,cos2\theta,...\}$. The the $[U_{pp}]$ component is $e^TU_pe=u-v$. Thus if there is only interpocket interaction it is $-v$. This explains the above structure for $[U]$. The matrix $[\tilde\Pi-\Pi]$ is
\beq\label{eq:matr2}
[\tilde\Pi-\Pi]=\left(\begin{array}{cccccc}
2[\mathcal{P}^{(1)}_h]&0&0&0&0&0\\
0&[\mathcal{P}^{(1)}_e]&0&0&0&0\\
0&0&[\mathcal{P}^{(2)}_h]&0&0&0\\
0&0&0&[\mathcal{P}^{(2)}_e]&0&0\\
0&0&0&0&[\mathcal{P}^{(3)}_h]&0\\
0&0&0&0&0&[\mathcal{P}^{(3)}_e]
\end{array}
\right),
\eeq
where $[\mathcal{P}^{(n)}_a]$ is a $2\times2$ matrix for the $n^{\rm th}$ harmonic channel for band $a$ and is of the form
\beq\label{eq:matr3}
[\mathcal{P}^{(n)}_a]=\left(\begin{array}{cc}
-\Pi^{nn,a}_{22}&-\Pi^{nn,a}_{23}\\
0&0
\end{array}
\right).
\eeq
The coefficient matrix $[c]$ is given by:
\beq\label{eq:matr4}
[c]=\left(\begin{array}{cc}
c^{(1,h)}_{2\times2}&0\\
0&c^{(1,e)}_{2\times2}\\
c^{(2,h)}_{2\times2}&0\\
0&c^{(2,e)}_{2\times2}\\
c^{(3,h)}_{2\times2}&0\\
0&c^{(3,e)}_{2\times2}
\end{array}
\right).
\eeq
Using the above expressions, it is easy to evaluate that
\beq\label{eq:Pirr}
\Pi^{RR}=\left(c^{(1,e)}\right)^2\left\{ \Pi^{11,e}_{33} - \frac{\left(\Pi^{11,e}_{23}\right)^2}{-\frac{2}{U^{(e)}}-\Pi^{11,e}_{22}} \right\} +
\left(c^{(2,h)}\right)^2\left\{ \Pi^{22,h}_{33} - \frac{\left(\Pi^{22,h}_{23}\right)^2}{\frac{2}{U^{(h)}}-\Pi^{22,h}_{22}} \right\} + \left(c^{(3,e)}\right)^2\Pi^{33,e}_{33}+\left(c^{(3,h)}\right)^2 \Pi^{33,h}_{33} ,
\eeq
where,
\bea\label{eq:fg}
&&\Pi^{22,h}_{22}=-\frac{2}{U_s}-2\nu_h \left(\frac{\Omega}{2\Delta}\right) ^2F(\Omega); ~~ \Pi^{22,h}_{33}=-2\nu_h F(\Omega);~~
\Pi^{22,h}_{23}=2\nu_h\frac{i\Omega}{2\Delta}F(\Omega),\nonumber\\
&&\Pi^{11,e}_{22}=-\frac{2}{U_s}-4\nu_e \left(\frac{\Omega}{2\Delta}\right) ^2F(\Omega); ~~ \Pi^{11,e}_{33}=-4\nu_e F(\Omega);~~
\Pi^{11,e}_{23}=4\nu_e\frac{i\Omega}{2\Delta}F(\Omega).
\eea
Here $F(\Omega)= \frac{\sin^{-1}\left(\Omega/2\Delta\right)}{\left(\Omega/2\Delta\right) \sqrt{1-\left(\Omega/2\Delta\right)^2}}$. The differencs between the hole and electron pocket expressions are due to the multiplicity of 2 for the electron pockets. It is easy to see for the case discussed in the main text that $\Pi^{nn,h/e}_{ij}=\Pi_{ij}$. To see the result in the main text, we relabel $c^{1,e},c^{2,h}\rightarrow c^e,c^h$ and $(c^{3,h})^2+(c^{3,e})^2\rightarrow(c^{o})^2$.

\subsection{BS modes in a two band model: A general case}
Here we wish to look at a case of BS modes in a two band model where the pockets are centered around the $\Gamma-$point. We will look at only the $\cos2\theta$ harmonic but include all inter and intra band interactions in the $d-$ and $s-$channels. Then using the general formula for the Raman response, we get
\bea\label{eq:new}
\Pi_{RR}&=&(c_1)^2\Pi^1_{33}+(c_2)^2\Pi^2_{33}+
\frac{(c_1)^2(\Pi^1_{23})^2\left[-\frac{U_{1d}}{2}+\Pi^2_{22}\frac{U_{1d}U_{2d}-V^2_d}{4}\right]+\left[1\leftrightarrow2\right]-2c_1c_2\Pi^1_{23}\Pi^2_{23}\frac{V_d}{2}}{\mathcal{D}},\\
\Pi^i_{22}&=&-2l_i-\left(\frac{\Omega}{2\Delta_i}\right)^2F_i,\\
\Pi^i_{23}&=&\left(\frac{i\Omega}{2\Delta_i}\right)F_i,\\
\Pi^i_{33}&=&-F_i,
\eea
where $\mathcal{D}=(1-U_{1d}\Pi^1_{22}/2)(1-U_{2d}\Pi^2_{22}/2)-\Pi^1_{22}\Pi^2_{22}V_d^2/4$. Further the ground state requires
\beq\label{eq:new2}
\left(\begin{array}{c}
\Delta_1\\
\Delta_2
\end{array}\right)=
-\left(\begin{array}{cc}
U_{1s}&V_{s}\\
V_s&U_{2s}
\end{array}\right)\left(\begin{array}{c}
\Delta_1l_1\\
\Delta_2l_2
\end{array}\right).
\eeq
For brevity, it will be useful to define
\bea\label{eq:new3}
&&\tilde{U}_{1s,2s}\equiv\frac{U_{1s,2s}}{U_{1s}U_{2s}-V_s^2},~\tilde{V}_{s}\equiv\frac{V_s}{U_{1s}U_{2s}-V_s^2};\\
&&\tilde{U}_{1d,2d}\equiv\frac{U_{1d,2d}}{U_{1d}U_{2d}-V_d^2},~\tilde{V}_{d}\equiv\frac{V_d}{U_{1d}U_{2d}-V_d^2};\\
\eea
Then,
\beq\label{eq:new6b}
\mathcal{D}=\left\{\left(\frac{\Omega}{2\Delta_2}\right)^2F_2+2(\tilde{U}_{1d}-\tilde{U}_{1s})+2\tilde V_s\frac{\Delta_1}{\Delta_2}\right\}
\left\{\left(\frac{\Omega}{2\Delta_1}\right)^2F_1+2(\tilde{U}_{2d}-\tilde{U}_{2s})+2\tilde V_s\frac{\Delta_2}{\Delta_1}\right\}-4\tilde V_d^2,
\eeq
This leads to $\Pi_{RR}=$
\beq\label{eq:new8}
\frac{-\mathcal{C}_{ds}(F_1c_1^2+F_2c_2^2)
-2F_1F_2\left[c_1^2(\tilde U_{2d}-\tilde U_{2s})\left(\frac{\Omega}{2\Delta_2}\right)^2+
c_2^2(\tilde U_{1d}-\tilde U_{1s})\left(\frac{\Omega}{2\Delta_1}\right)^2\right]
-2F_1F_2\left(\frac{\Omega}{2\Delta_1}\right)\left(\frac{\Omega}{2\Delta_2}\right)
\left[(c_1^2+c_2^2)\tilde{V}_s-2c_1c_2\tilde V_d\right]}{\mathcal{D}},
\eeq
with
\beq\label{eq:new6a}
\mathcal{C}_{ds}=\left\{2(\tilde{U}_{1d}-\tilde{U}_{1s})+2\tilde V_s\frac{\Delta_1}{\Delta_2}\right\}
\left\{2(\tilde{U}_{2d}-\tilde{U}_{2s})+2\tilde V_s\frac{\Delta_2}{\Delta_1}\right\}-4\tilde V_d^2.
\eeq

There are several things to note here:
\begin{itemize}
\item sgn$(\Delta_1\Delta_2V_s)=-1$. Thus with the exception of the lone $\tilde V_d$ term in the numerator of $\Pi_{RR}$, there is no sensitivity to $s\pm$ vs $s++$ (which is decided by the sign of $V_s$).
\item If we set all the $d-$wave interactions equal to $s-$wave interactions, then we arrive at the ``Leggett mode scenario''.
\item The BS mode loses weight as it softens. In fact, for 1band case, the Raman response is zero at the softening of the BS mode (the $s+id$ transition).
\end{itemize}

\subsubsection{The role of interband $d-$wave interaction}
Here we first switch off the interband $d-$wave component $V_d$. The above expression for the Raman response reduces to
\bea\label{eq:new9}
\Pi_{RR}&=&-F_1c_1^2\frac{2(\tilde U_{2d}-\tilde U_{2s})+2\tilde V_s\frac{\Delta_2}{\Delta_1}}{\left(\frac{\Omega}{2\Delta_1}\right)^2F_1+2(\tilde U_{2d}-\tilde U_{2s})+2\tilde V_s\frac{\Delta_2}{\Delta_1}}-F_2c_2^2\frac{2(\tilde U_{1d}-\tilde U_{1s})+2\tilde V_s\frac{\Delta_1}{\Delta_2}}{\left(\frac{\Omega}{2\Delta_2}\right)^2F_2+2(\tilde U_{1d}-\tilde U_{1s})+2\tilde V_s\frac{\Delta_1}{\Delta_2}}
\eea

Anlayzing Eq. \ref{eq:new9} tells us
\begin{itemize}
\item If $V_d=0$ (in the presence of $U_{1d,2d}$ interactions), each pole corresponds to a potential BS mode (depends on the interaction). Each solution in $T_c$ will give a BS mode.
\item $-\tilde U_{2s}+\tilde V_s\Delta_2/\Delta_1 = l_1$ and $-\tilde U_{1s}+\tilde V_s\Delta_1/\Delta_2 = l_1>0$. So the mode frequencies are given by(Matsubara frequencies) $\Omega_1^2=-2\tilde U_{2d}-2l_1=-2/U_{2d}-2l_1$. We see that the parameters we need to get the BS modes are $T=T_c$ for $d-$wave and $T=0$ for $s-$wave.
\item No distinction can be drawn between $s\pm$ vs $s++$.
\end{itemize}

We now introduce the interband $d-$wave interaction. Lets introduce the abbreviations:
\bea\label{eq:new10}
&&A_1\equiv2(\tilde U_{2d}-\tilde U_{2s})+2\tilde V_s\frac{\Delta_2}{\Delta_1};
~~A_2\equiv2(\tilde U_{1d}-\tilde U_{1s})+2\tilde V_s\frac{\Delta_1}{\Delta_2};\nonumber\\
&&\Omega_i^2\equiv\left(\frac{\Omega}{2\Delta_1}\right)^2F_i,~~i\in{1,2}.
\eea
Lets re-write the response in the condensed notation:
\bea\label{eq:new11}
\Pi_{RR}&=&\frac{-F_1c_1^2 A_1(\Omega_2^2+A_2)-F_2c_2^2A_2(\Omega_1^2+A_1)+4\tilde V_d^2(F_1c_1^2+F_2c_2^2)+4F_1F_2c_1c_2\tilde V_d\left(\frac{\Omega}{2\Delta_1}\right)\left(\frac{\Omega}{2\Delta_2}\right)}
{(\Omega_1^2+A_1)(\Omega_2^2+A_2)-4\tilde V_d^2}
\eea
\begin{itemize}
\item The only term sensitive to $s\pm$ vs $s++$ is in the spectral weight in the linear in $\tilde V_d$ term.
\item Introducing $V_d$ causes damping of the mode between $2\Delta_1$ and $2\Delta_2$. Thus there can be two true collective modes, or only one true collective mode and one resonance per attractive subdominant channel.
\item Introducing $V_d$ causes level repulsion between the modes. As it is increased, one mode is pushed towards zero and the other towards the $2\Delta$ of the larger gap.
\item Increasing $V_d$, one solution to $T_c$-equation vanishes when $V_d^2=U_{1d}U_{2d}$.  But before the second solution is lost, the leading solution softens.
\item Thus if $U_{1d}=0=U_{2d}$, then we seem to get 1 mode and possibly one resonance.
\item It is easy to see that a BS mode will soften when $A_1A_2=4\tilde V_d^2$. {And the weight of the BS mode goes to zero.}
\end{itemize}

To see level repulsion: write $2F_1F_2\left(\frac{\Omega}{2\Delta_1}\right)\left(\frac{\Omega}{2\Delta_2}\right)=
\frac{\Delta_2}{\Delta_1}\Omega_2^2F_1+\frac{\Delta_1}{\Delta_2}\Omega_1^2F_2$. Then find correction to $\Omega$ to leading order in $V_d$. This means $\Omega_1^2\rightarrow -A_1+ f_1\tilde V_d^2$ and $\Omega_2^2\rightarrow -A_2+ f_2\tilde V_d^2$. This immediately yields (we assume for definiteness: $\Delta_1^2<\Delta_2^2\rightarrow |A_1|<|A_2|$),
\bea\label{eq:new12}
f_1&=&\frac{4}{\Omega_2^2|_{\Omega_{BS,1}}+A_2}
\approx\frac{4}{-\left(\frac{\Delta_1}{\Delta_2}\right)^2A_1+A_2}<0,\nonumber\\
f_2&=&\frac{4}{\Omega_1^2|_{\Omega_{BS,2}}+A_1}
\approx\frac{4}{-\left(\frac{\Delta_2}{\Delta_1}\right)^2A_2+A_1}>0.
\eea
The two different signs indicate that the modes repel each other. In general, the lower mode softens and the upper mode goes towards the respective $2\Delta$. Softening seems to happen before one of the solutions expire.

\subsection{RPA calculations for Ba$_{1-x}$K$_x$Fe$_2$As$_2$}

The results of the RPA calculations presented in Figs.~4c and d were obtained for the 5-orbital Hubbard-Hund model for BaFe$_2$As$_2$ introduced by Graser {\it et al.} in Ref.~\cite{Graser10}. While the electron densities were obtained in the 3D model, we used a 2D $k_z=0$ cut for the pairing calculations. The pairing eigenvalues and eigenvectors  were obtained using the RPA formalism presented in Ref.~\cite{Graser10}. In this framework, one first calculates the RPA enhanced spin and charge susceptibilities, which then enter the pairing interaction in the usual fluctuation exchange approximation \cite{bickers89,takimoto04}. This interaction is then used in the linearized BCS gap equation restricted to the vicinity of the Fermi surface to determine the pairing strengths (eigenvalues) for the $s$-wave state, $\lambda_s$ and for the $d$-wave states, $\lambda_d$, and their corresponding eigenvectors. For the local Hubbard interactions we have used $U=0.90$ eV (intra-orbital Coulomb) and $U'=U/2$ (inter-orbital Coulomb) and set the Hund's rule coupling ($J$) and pair-hopping ($J'$) interactions to $J=J'=U/4$ satisfying spin-rotational invariance.
\end{widetext}

\bibliographystyle{apsrev4-1}

\begin{thebibliography}{10}
\bibitem{FN} In order to get a B$_{\rm 1g}$ response, an anisotropic dispersion is required;   however, for simplicity, we assume (in calculation of the $\Pi$s) that the dispersion can be linearized near the FS and can be assumed to be isotropic, with all information about anisotropy absorbed in the overall angular factors.
\bibitem{FN2} There is a B$_{\rm 2g}$ solution too, but this does not couple to the B$_{\rm 1g}$ probe and hence we do not discuss it.
\end{thebibliography}

\end{document}